# Enacting planets to learn physics

**E ROLLINDE[1], N DECAMP[2]**

[1] *Laboratoire de Didactique André Revuz, EA 4434, Université Cergy-Pontoise, 33, boulevard du Port, France*
[2] *Laboratoire de Didactique André Revuz, EA 4434, Université Paris Diderot, 5 rue Thomas Mann, 75013 Paris, France*

**Abstract.** The Solar System motivates students to interest themselves in sciences, as a large number of concepts may be easily introduced through the observation and understanding of planet's motion. Using a large representation of the Solar System at a human scale ("a human Orrery"), we intend to show how cognitive activities about kinematics and dynamics are activated and linked to the sensori-motors activities. In the last three years, we have conducted different activities with 10 to 16 years old children. In this contribution, we discuss the different scientific concepts covered by the Human Orrery, and the enaction theory that provides a theoretical background to those activities. We detail the enaction of velocity with a description of the gestures in relation with the abstract concepts involved in kinematics.

## 1. Introduction

In 2013, a program called "kinaesthetic activities in teaching science and humanities" was granted by Sorbonne University, France, connecting UPMC (departments of physics and biology), and Paris Sorbonne (departments of sports, Italian, and ancient Greek). A "human Orrery" was thus created (Figure 1, [1]), allowing the learners to enact the planets' movement with correct relative speed. An Orrery is a mechanical or digital device designed to illustrate the motion of the planets around the Sun and their changing positions in the sky. On a human Orrery, the orbits of planets and comets are drawn at a human scale allowing movements in the Solar System to be enacted by the learners. The implementation of Human Orrery, as learning experience, involves topics from mathematics and science and illustrates an example of a STEM approach of learning about concepts perceived as abstracts by students and teachers. Astronomy provides a highly motivating context for learners to develop observational skills, discover methods of scientific inquiry, and explore some of the fundamental laws of physics and concepts of mathematics in both an attractive and meaningful way.

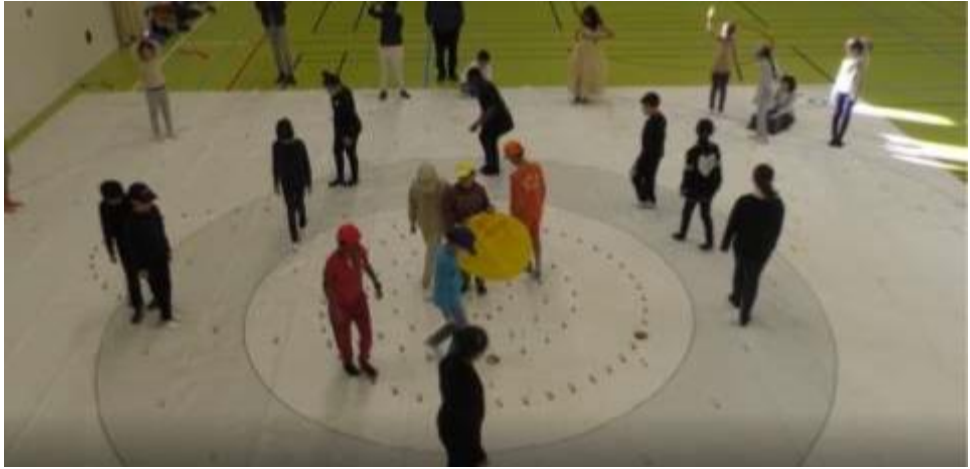

**Figure 1.** The "Human Orrery": Pupils enact planets (Mercury to Jupiter), asteroids and comets (in black) along their orbits around the Sun

*1.1 How to use a Human Orrery*

The design of a Human Orrery is made such that users may walk along orbits of different bodies around the Sun that is located at the centre of the design. The Human Orrery that we use is printed in a large map of 12m by 12m (see Figure 1). It allows one to follow the orbits of the inner planets (Mercury, Venus, Earth and Mars) and Jupiter; the inner planets are located inside the asteroid belt that is materialized with a grey colour together with the orbit of the largest asteroid known, Cérès. The highly eccentric orbits of two comets are also used: Encke (the smallest elliptical orbit) and Chury. This choice of objects illustrates different type of movements while keeping the size of the Orrery reasonable. Earth is at one meter from the Sun (spatial scale), while Jupiter's orbit has a diameter of 10,5m. The orbits of all bodies are materialized by dots at constant intervals of times, with accurate elliptical shapes. Note that orbits are near-circular for the five planets and Ceres.
 The interval of time may be different for each orbit, but is always a multiple of 16 terrestrial days: For Earth, there are 23 dots separated by 16 days, which would make a period of 368 days. For Jupiter, there are 54 dots separated by 80 days, which makes a period of 4320 days instead of the real period of 4332,59 days. Note that this difference may be used as an introduction to the Euclidian division in mathematics. A sound (either a clock or hand claps) is heard regularly. The interval of time between two sounds corresponds to 16 terrestrial days (temporal scale). Every user makes one step during this time interval. Then, the person that enacts Earth walks from one point to the next (distance) in one step (duration), while "Jupiter" has to do five steps (five times 16 days) to reach the next point. All rules are described in details in [2]. By acting according to those rules, the movements on the Human Orrery illustrate the correct relative velocities of all Solar System bodies.

*1.2 Scientific concepts in use*

The topics of different sequences tested since 2015 are described briefly here. (i) Construction of a human Orrery. The description of the Solar System involves different length scales from the diameter of the planets to the distance between planets and the length of one orbit. It also involves different duration scales from the rotation to the orbital period. The construction of a Solar System implies a choice of length scales and orientation and the drawing of ellipses. The links with mathematics is obvious, including placement on a 2D plan, Euclidian division and simple geometry. (ii) Enacting planets to learn physics. In the last 3 years, planets have been enacted by undergraduate students in Paris as well as pupils of primary and secondary schools in Paris, Nantes, and Beirut on their own Orrery or on Sorbonne University one. Through the eccentric orbit of a comet, students have observed its varying velocity and realized why force and speed are not directly related, what is the meaning of the work-energy theorem… 16-year-old students have observed, measured and plot Kepler's laws… 12-year-old pupils have experienced that a larger distance may be travelled in a larger amount of time if the velocity is smaller… and 10-year-old pupils have made a movie about shooting stars (encounter of Earth and a comet during the night).

Topics such as inertial movement, velocity-duration-distance and force-velocity relations, known to be difficult, can be refined and perceived by the learners' body. The Orrery has been used in France and in Lebanon in different pedagogical contexts with 10 to 16 years old pupils [2]. This empirical work is based on the cognitive science theory of enaction [3] that is already well known and widely used in Science and Mathematics Education [e.g. 4; 5; 6; 7].

## 2. Enaction theory as a theoretical background

The use of a Human Orrery in education is based on the assumption that bodily perceptions help the learning of abstract concepts. The conversion of learning space into performance space means that embodiment becomes a vehicle for interpretation [8], and movement a medium for choreographing thought [9]. New connections are set up between mental activity and kinetic activity; sense-making and sensory-motor experience. Such an assumption is based on the cognition theory of enaction [3].

For further theoretical details, we refer the reader in particular to the analysis of the theory of enaction proposed by [10]. Traditional psychological thinking in the line of Plato, considers the body as an obstacle to thought. The epistemology of Piaget [11] extends this vision by considering the sensorimotor stage as a transition to abstract thought and distinguishing between sensory experiences and cognition. With a didactic point of view (see for example, [12] and [13]) the daily sensory experience is seen as the source of a first learning (or a first questioning) but it can also become a brake on the understanding of phenomena because it involves factors that we are not aware of (presence of friction for example). The setting up of well-defined abstract concepts is then limited by too strong associations with these global perceptions. Enaction theory considers the interaction between the body and the outside world as the foundation of cognition, what Lapaire [9] calls "kineflexion" and Radford [14] "a sensuous cognition". Abstract concepts must then be embodied and experienced in sensory experiences to be fully integrated [15]; the environment becomes then an actor of reflection under the condition that it favours a fine perception [3]. This distinction between fine perception and coarse perception proposed by Varela brings us closer to the conclusions of didactics.

## 3. A gestural dictionary of enacting velocity: an exemplar illustration of congruency

To be in line with the enaction theory, as shortly described above, the use of the Human Orrery has to provide students with an adapted or congruent sensory pathway to the abstract notions to be learned (see [16] and [17]). We describe here in more detail how gestures and movements on the Orrery are associated with abstract concepts of kinematics. The following can be read as a first "dictionary of the Human Orrery". The long-term goal is to foster "an interaction between different sensory modalities and different signs" that will allow learners "to perceive a theoretical dimension" in their "associated mathematical perceptions and thoughts" [10].

*3.1 Different walks along the orbits*

To make learner walk along the orbit of the planet, the game master (here the teacher, but it could be a student also) claps hands regularly. The learner (or walker) has to stay on one orbit by following the associated points on the Orrery (Figure 2). At each clap of the leader of the game, the learner must hence take a step following the orbit. The teacher can change the frequency of claps and the learner may change the length of the step. By jumping one or more points at each step (Figure 2, centre), the distance travelled over the same duration increases and therefore the speed increases. On the contrary, by taking several steps between two points (Figure 2, right), the distance travelled over the same duration decreases, as well as the speed. Finally, by changing the duration between two claps, the duration of a step varies, as well as the speed.

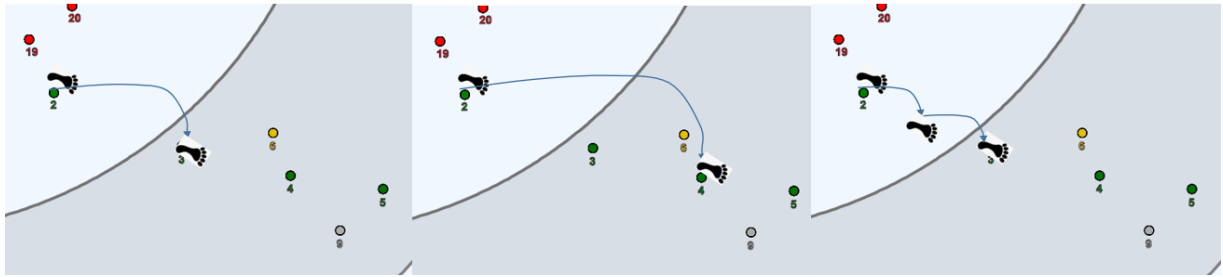

**Figure 2.** Different walk along the orbit of the comet Encke (green points). At each step, one may walk to the next point (left), to the second next point (centre) or one may take two steps to reach the next point (right).

If one follows the rules given in Section 1 for each celestial body on the Human Orrery, the relative velocities during the choreography are those of the real planets. We now detail the association between gestures and words.

*3.2 Gestures congruent to the notions of instant and duration*

The action of the game master (clapping his hands) sets the "musical" rhythm. The distinction between instant and duration is a very important one while learning kinematics and must be clearly apparent in the words used for the description as well as in the mind of the learner doing the choreography.

*A clap*: At the moment the game master claps his hands, a sound is heard. A "clap" may have different meaning; the perception of the hands that are struck (for the master of the game), the audition of the sound (for the actors) and the temporal instant (for all).

*The duration between two claps or the frequency of the claps*: The duration between two claps is constant (according to the regularity of the master of the game ...). We have noticed that both the students and the teacher often speak of "frequency", sometimes implicitly, that is, the number of claps heard in a second (although this is never explicitly stated). It may be better to speak only of the duration between two claps and to define the notion of frequency, whose relationship with the speed is more complex, later. Note that the duration of a clap is meaningless because a clap is considered as an instant.

*The duration of a step*: this duration is imposed as equal to the duration between two claps, while the length of a step is left free. We will return to this difference in the analysis of the embodiment of speed.

*The duration of a revolution around the Sun*: this is a particular time that can be problematic. The knowledge of the duration of a revolution for a planet sets its angular velocity and also its linear velocity because the distance travelled is fixed. In the case of the movement of two bodies (Earth and Mercury for example); if they finish one turn simultaneously, they would have the same angular velocity but different linear velocities. Although angular velocity is not part of the "college" curriculum (in France), this notion was intuitive for 12 years-old students and implicitly present in their speech.

*Duration scale:* The interval of time between two claps corresponds to the real duration of 16 terrestrial days. This provides an exact scale. Another, more intuitive, way to define the scale relates to the terrestrial year. If one measures the duration of one turn of the learner embodying the Earth, it corresponds then to a "real" duration of one year. For younger learners, this is certainly a better way to enact the idea of duration scale.

*3.3 Gestures congruent to the displacements*

Three types of displacements are possible on the Orrery. Again, the association of these displacements with positions and distances must be clearly specified. The position / distance (or length) distinction is again fundamental to understand the concept of speed, as well as its link with instant/duration.

*One step*: "One step" is the movement made during the interval of time between two claps. This displacement does not necessarily go from one point of the planet's orbit to another. Thus, the initial and final positions of a "step" are not imposed by the Orrery design. Hence, the duration of this displacement is imposed, but not its length.

*The points of the Orrery*: Each point corresponds naturally to a position since the displacements of the pupils are made in order to go from one point to another. This can be done in several steps, but each point of the Orrery will be reached at a given moment. The distance between the points is observed directly on the Orrery and will be travelled in one or more steps. Hence, the duration of the movement from one point to another is not imposed, but its length is fixed.

*The perimeter of an orbit*: This is a particular distance on the Human Orrery. Again, the duration to go through it is not imposed (it depends on the temporal scale or on the chosen frequency of the claps), but its length is fixed.

*Distance scale*: This scale is imposed by the Orrery design and cannot be modified. The half-major axis of the Earth can be determined geometrically in the drawing. This length measured on the Orrery corresponds to an "Astronomical Unit" for the real Solar System.

As we have seen in the two previous sections, displacements (duration) do not have the same status as points (and instants) on the Human Orrery. These differences must be explained by the gesture or the word so as not to be confusing. The possible interplay between these notions then allows a fruitful questioning on the speed.

*3.4 How is the speed enacted?*

For an abstract concept to be enacted means that meaning has emerged from the interaction with our sensory-motor system [18]. In the context of kinematics, the length travelled is enacted by the vision (Orrery points) and the kinaesthetic action (displacement of the feet), in an active form. The duration is enacted by the hearing of the clap, in a passive form. How is the speed enacted?

The concept of speed is not visible (like distance) nor audible (like duration). In principle, no sense can "measure" a speed. Yet, everyone can feel, perceive if he has a great speed or not, if he slows down or accelerates. As a passenger in a car, looking at houses in the landscape, one knows that the speed of the car has increased as the houses, *a constant distance*, go past us in a shorter interval of time. The perception of the speeds of two cars is done by looking at the distance they cover for *a similar short duration*. Thus, the known didactic difficulties of handling a two-parameter relation (speed as a function of distance *and* duration) is never enacted in our everyday actions. Using the human Orrery allows to enact speed in connection with distance and duration (see [19] for a specific sequence that allows learners to enact the concept of proportionality).

To explain how the speed is enacted on the Orrery in relation to the notions of distance and duration, the movement of the feet during two steps and around 4 positions (A, B, C and D) is detailed (Table 1). Let us insist that these positions are not necessarily the points of the orbit (for example, the points on Encke's orbit would be located at positions A and D since 3 steps are needed to go from one point to the next one).

**Table 1.** Description of the choreography during two steps and three claps.

| Instant 1 | First step | Instant 2 | Second step | Instant 3 |
|---|---|---|---|---|
| Right foot is on position A, left | Right foot goes from position A to C. The whole body has moved along a | Right foot is on position C, left foot on B. A | Left foot goes from position B | Right foot is on position C, left foot on D. A |

| foot on B. A clap is heard. | given *distance* during a given *interval of time*. | second clap is heard. | to D. It reaches D at the Instant 3. | third clap is heard. |

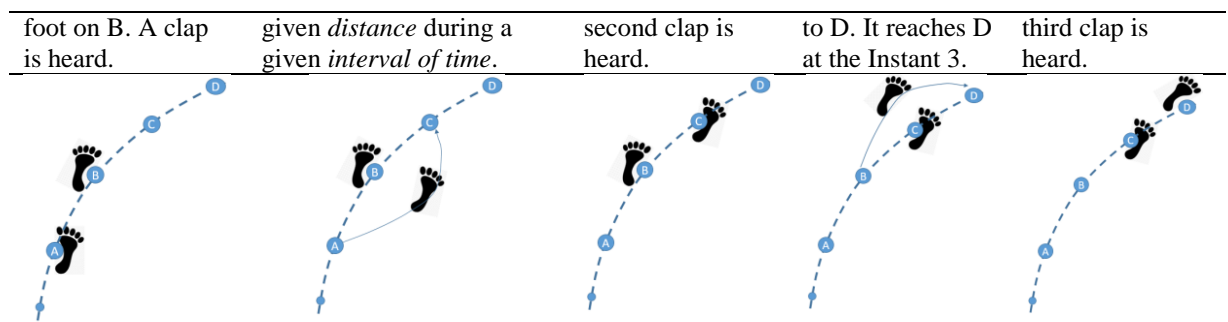

This specific choreography can be modified to enhance sensory-motors perceptions, while remaining attentive to two important aspects. Firstly, the distinction between time / duration and position / displacement must be clear to the teacher in order to bring these concepts to the student without confusion. Secondly, the movement must be done without halts: avoid resting at one position between two claps and moving quickly at the instant of the clap. This is necessary so that the notion of instantaneous speed has a meaning in this choreography.

From these elements of language associated with the gestures, the use of human Orrery can lead to an effective implementation of the notions of kinematics.

**4. Discussion**

Through our initiative, seven Human Orreries were built in France (one for a science centre, one in a public place in Paris – Le Jardin des Plantes, five in primary or secondary schools), one was drawn in a Lebanon school and one map was purchased by a science centre in Vietnam. Human Orreries are used thereafter by different teachers (physics, mathematics, technology) in those places. Different sequences have been co-constructed by teachers and researchers. Among those, three experimentations have led to deeper analysis or nice outputs: (i) We have revealed an enhanced understanding of kinematics concepts through an experiment conducted during three years with 16-year-old students [2]; (ii) a movie realised by 10-year-old pupils that explains the origin of shooting stars, enacting thus different referential frames without naming them, is available on-line (planetaire.overblog.com); we announce here that in 2018-2019, a larger experimentation on referential frames will be set up with 16-year-old learners; (iii) in 2017-2018, we conducted a large experimentation on speed with four classes of 12-year-old pupils in a French school. The detailed sequence will be published elsewhere and may be requested directly to the author. One of the most striking observation during those sequences was a global confusion in the use of kinematic words and the difficulty for the teacher to have the word "duration" emerge in the learner's talk. Hence our idea to create the "dictionary" proposed in this contribution.

In recent years, research on learning and education has been increasingly influenced by theories of embodied cognition. Several embodiment-based interventions have been empirically investigated, including gesturing, interactive digital media, and bodily activity in general (see [7] for a review). The research presented here is in line with this field of research. The setting-up of our design bring us to the highest level of embodiment as defined by [17] and refined by [7], as it is using *integrated forms of embodied learning* (hence the dictionary) and a high level of bodily engagement. Although our design does not rely on virtual reality, the immersion of learners into the environment is certainly high. However, the risk of cognitive overload [20] may explain the marginal effect on learning found in [2]. Further research towards evaluation of those sequences in terms of learning improvement and measurement of cognitive loads [21] will be considered in the context of a larger project as described next.

**5. Conclusion and Perspectives**

The perspective that we adopt in our project is that of STEM being taught through an interdisciplinary approach that incorporates an embodied dimension and draws on real-world modelling. The use of the Human Orrery enables students to enact or "experience" scientific concepts and the dynamics of their properties by evolving in an adapted environment. Until recently, we only focused on the dissemination of Human Orreries and on a proof-of-concepts in terms of motivation and interest. In this contribution, we have detailed the use of the human Orrery and the integrated form of embodied learning used in the context of kinematics.

The current dynamic of the project is to motivate more classes from different levels and teachers from subjects other than mathematics and physics such as art, technology, music, languages (native or foreign), physical education and sport. In 2018-2019, more schools have decided to join the project in France: two primary schools in the context of interdisciplinary projects and one secondary school through a project involving science and technology. At the European level, we are expecting to submit a H2020 project involving research in mathematics and science didactics together with an open schooling perspective (museum, observatories and associations). Our next objective is then to design sequences that fit into the STEAM approach and that will be used as experimental situations for the theoretical framework of enaction, complemented with the framework of Activity Theory in the context of science education [22]. Our focus will be as much on the didactical learning as on the motivation and awareness of the learners during the activity.